\documentclass[10pt]{wlscirep}
\usepackage[utf8]{inputenc}
\usepackage[T1]{fontenc}
\title{The Fallibility of Contact-Tracing Apps}

\author[1,*]{Piotr Sapiezynski}
\author[2]{Johanna Pruessing}
\author[3,4]{Vedran Sekara}
\affil[1]{Khoury College of Computer Sciences, Northeastern University, Boston, USA}
\affil[2]{Open Society Foundations, Berlin, Germany}
\affil[3]{IT University, Copenhagen, Denmark}
\affil[4]{UNICEF, New York, USA}

\affil[*]{corresponding author: p.sapiezynski@northeastern.edu}
\begin{abstract}
Since the onset of the COVID-19's global spread we have been following the debate around contact tracing apps--the tech-enabled response to the pandemic. As corporations, academics, governments, and civil society discuss the \textit{right} way to implement these apps, we noticed recurring implicit assumptions. The proposed solutions are designed for a world where Internet access and smartphone ownership are a given, people are willing and able to install these apps, and those who receive notifications about  potential exposure to the virus have access to testing and can isolate safely. In this work we challenge these assumptions. We not only show that there are not enough smartphones worldwide to reach required adoption thresholds but also highlight a broad  lack of internet access, which affects  certain groups more: the elderly, those with lower incomes, and those with limited ability to socially distance. Unfortunately, these are also the groups that are at the highest risks from COVID-19.
We also report that the contact tracing apps that are already deployed on an opt-in basis show disappointing adoption levels. We warn about the potential consequences of over-extending the existing state and corporate surveillance powers. Finally, we describe a multitude of scenarios where contact tracing apps will not help regardless of access or policy. In this work we call for a comprehensive and equitable policy response that prioritizes the needs of the most vulnerable, protects human rights, and considers long term impact instead of focusing on technology-first fixes.

\end{abstract}
\begin{document}

\flushbottom
\maketitle
\thispagestyle{empty}

\section*{Introduction}

Communicable diseases, such as COVID-19, spread between people in close proximity. This is why, in addition to social distancing, business closures, and wearing masks, public health interventions include contact tracing. It is a method of identifying people that were in physical proximity with a positive patient, so that they can be isolated, tested and, if necessary, treated. Traditionally, contact tracing has been done by human tracers, who interview infected patients about all their recent contacts and then follow up with them. Human tracers engage directly with their  interviewees and are trained to address any potential  health and confidentiality concerns.~\cite{vogelstein2020} However, this approach has two drawbacks.
First, \textit{low recall}: we humans tend not to remember all the people we were in close contact with during the period where we were potentially infectious (up to two weeks before onset of symptoms~\cite{who2020}). It is also impossible to identify strangers we have been close to on the bus or in the store, etc.
And second, \textit{slow process}: human tracers have to reach out to all the reported contacts and talk to them. This can be a painstakingly slow process and it might well take days to get a hold of all the people we listed.
Smartphone based contact tracing promises to solve both of these limitations. If it works as intended we would not have to rely on our faulty memory. The app could ``remember'' the people we do not even know, as long as both we all have compatible smartphones and installed the app. Once we are diagnosed, the app would instantly notify all our contacts who meet certain criteria, such as  duration of contact, closeness, etc. and recommend testing and/or isolation, But of course only if everyone has a smartphone, the app, and reliable internet access.

Yet, actually solving these two drawbacks brings about a variety of technical challenges. Currently, we are seeing a growing consensus around a decentralized and privacy-preserving design using Bluetooth Low Energy (BLE) to measure proximity. One such approach was proposed by a consortium of European academics and later adopted by Google and Apple.~\cite{goodin2020} The few countries that chose more privacy-invasive approaches, including the UK, Norway, Australia, and Singapore quickly learned that their apps do not work on iPhones. As a result  some of them have signaled willingness to reconsider their approach. 

Bluetooth Low Energy was chosen instead of popularly discussed location data (from GPS, cell towers, or WiFi), because of its accuracy and battery efficiency. Location data is not accurate inside of buildings and therefore cannot be used to determine whether two people were actually close enough to exchange the virus, or just  happened to be in the same building. A location-based approach would lead to plenty of false  exposure notifications, by for example reporting everyone living in the same building as an infected person. For more information about how exactly the Bluetooth-based, decentralized tracing would work, we recommend a great infographic~\cite{case2020} by the Decentralized Privacy-Preserving Proximity Tracing group (DP3T), one of the groups working on implementing tracing apps.

\subsection*{Contribution}
Regardless of the architecture and particularities of implementation, the proponents of contact tracing apps appear to make a number of assumptions, which we challenge in this work:
\begin{itemize}
\item Do people have modern smartphones and unhindered access to the Internet? 
\item Can we use Bluetooth signals to measure physical proximity between people? Can this be actually epidemiologically useful?
\item Can and will individuals \textit{choose} to install contact tracing apps? 
\item Can individuals access tests and adapt their behavior accordingly once the app signals exposure?
\end{itemize}
When we answer these questions using the available data, we find that smartphone-based solutions to the pandemic may have vastly differential impacts, if deployed without considering local contexts and varying needs of different communities. The virus, initially dubbed by some a ``great equalizer''\cite{cuomo2020} has progressively proven to be  a ``great magnifier''\cite{diva2020} as it already disproportionately affected the most disadvantaged communities. It is paramount we ensure that any attempted technological interventions to the pandemic do not exacerbate existing inequalities and are, instead, a part of an equitable public health response. While it might appear that smartphone based contact tracing is a promising approach in selected geographies, we must not forget that the pandemic is, by the very nature of how much we are connected, a global problem and not one of separate societies. In the sections below we take a closer look at each of these implicit assumptions.

To be clear, this document is not a statement against ethically responsible, privacy-preserving, and time-limited contact tracing as a concept. Community based contact tracing, traditional interview and follow-up in person or via phone-call with those potentially exposed, and, in places where we can ensure viability, app-assisted tracing can be necessary elements of reopening countries once testing is widely available and equally accessible to everyone.

\begin{figure}[t]
    \centering
    \includegraphics[width=1\linewidth]{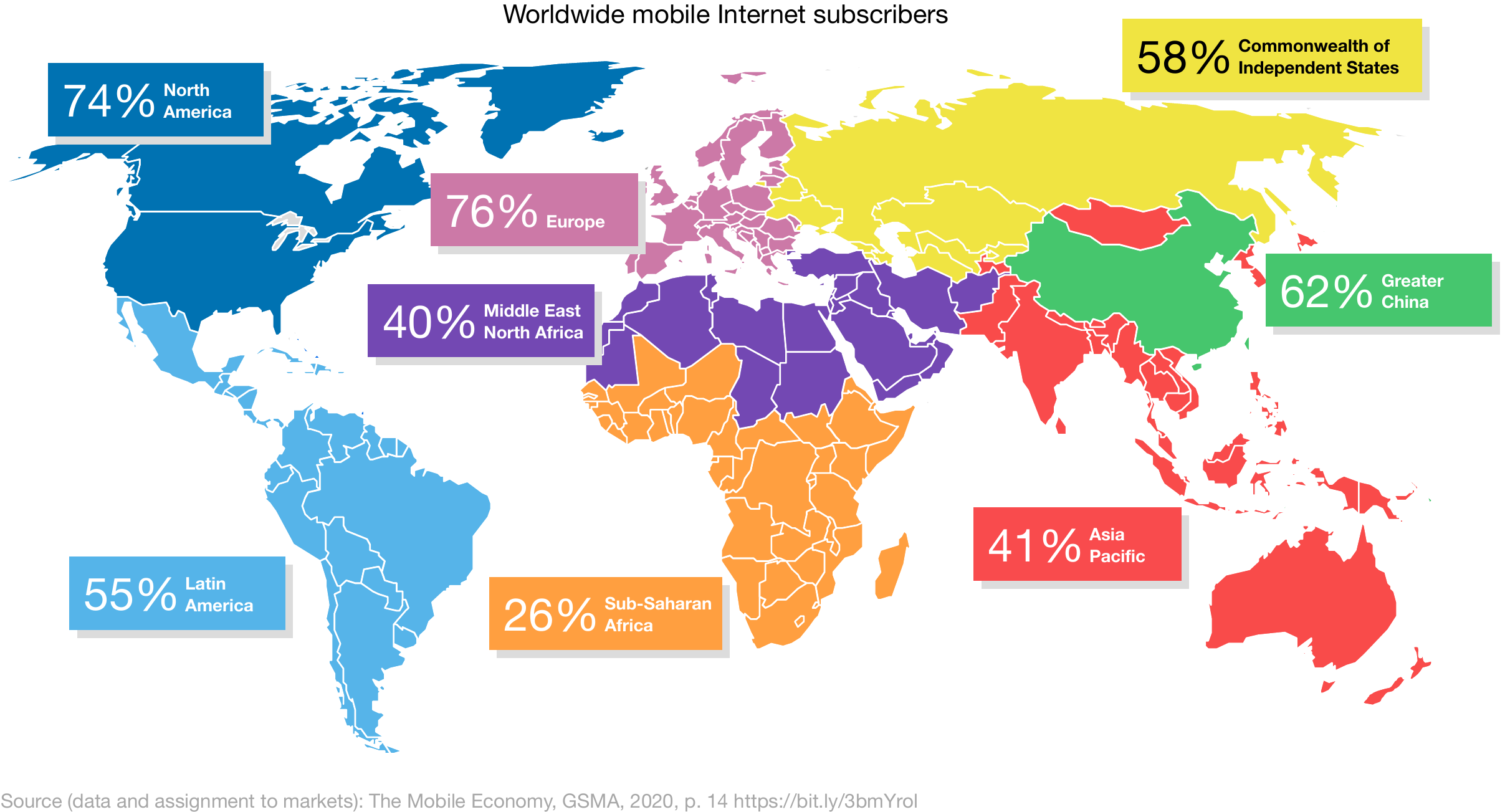}
    \caption{Access to mobile Internet is not evenly distributed across the world and is especially scarce in Sub-Saharan Africa.}
    \label{fig:gsma_map}
\end{figure}

\section*{Assumption 1: People have modern smartphones and Internet access}
While in many countries smartphones are taken for granted, worldwide, only approximately 42\%-45\% of the population has a smartphone~\cite{odea2020} and only about three quarters of these devices support the required Bluetooth Low Energy technology.~\cite{bradshaw2020} This means that more than half the world's population live in countries where smartphone ownership might not be high enough for smartphone-based contact tracing to work as we show below. The latest GSMA report (see Figure~\ref{fig:gsma_map}) further illustrates large differences between regions: in Sub-Saharan Africa only 26 out of 100 people have a mobile Internet subscription; for comparison, in Europe the number is 76 out of 100. Internet connectivity (necessary for contact tracing apps to work) is similarly unevenly distributed, leaving nearly half of the world's population, 3.5 billion people, without access.~\cite{itu2020} 

\begin{figure}[tb]
    \centering
    \includegraphics[width=0.75\linewidth]{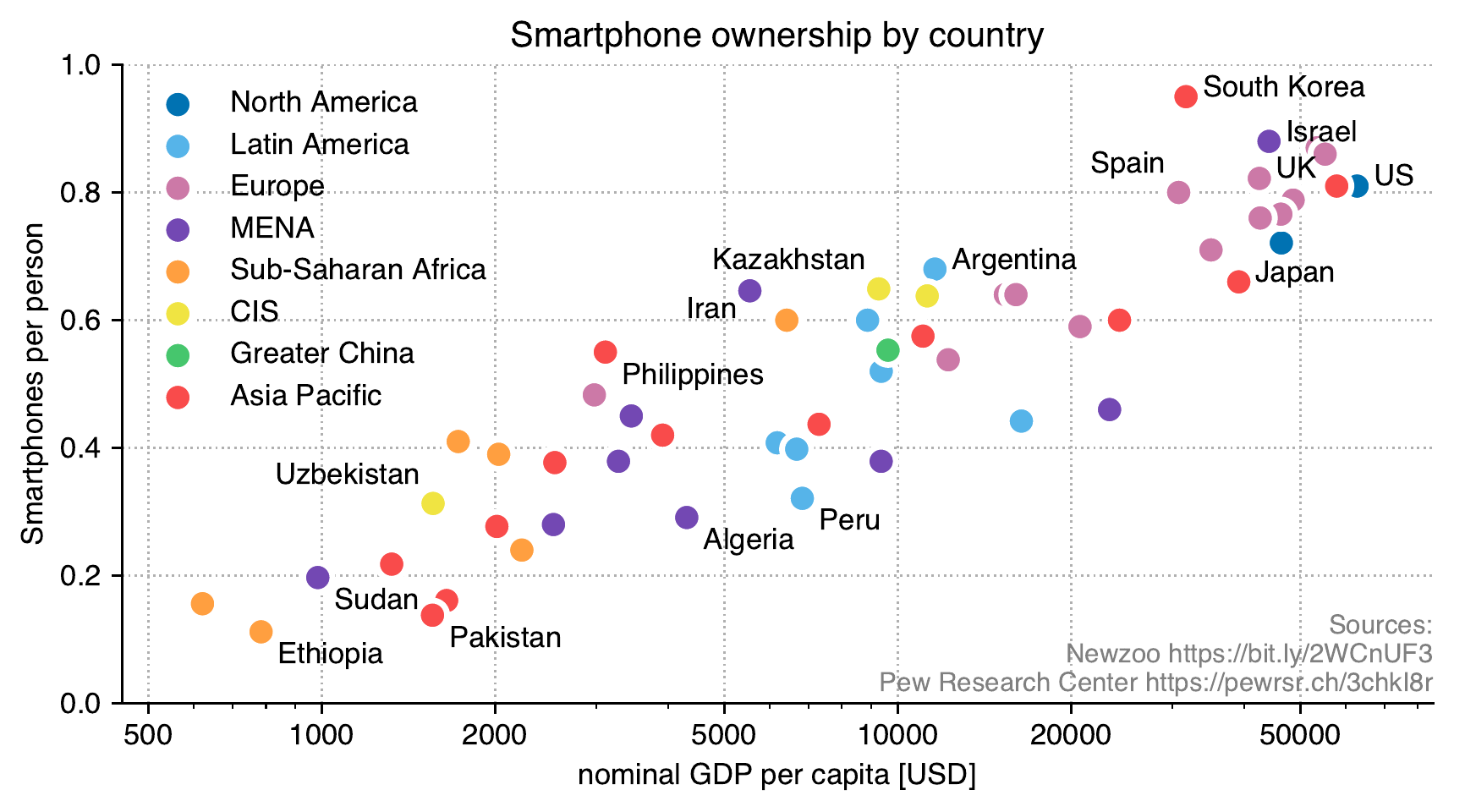}
    \caption{Smartphone penetration is strongly correlated with the wealth of countries--the richest countries have eight times more smartphones per capita than the poorest.}
    \label{fig:ownership}
\end{figure}

These staggering differences are related to the wealth of countries measured here in GDP per capita, see Figure~\ref{fig:ownership}. In the poorest countries  only one out of 10 people have a smartphone. This number is eight to nine times higher in the wealthiest countries meaning only one or two out of 10 people do not have one. 

\begin{figure}[tb]
    \centering
    \includegraphics[width=0.75\linewidth]{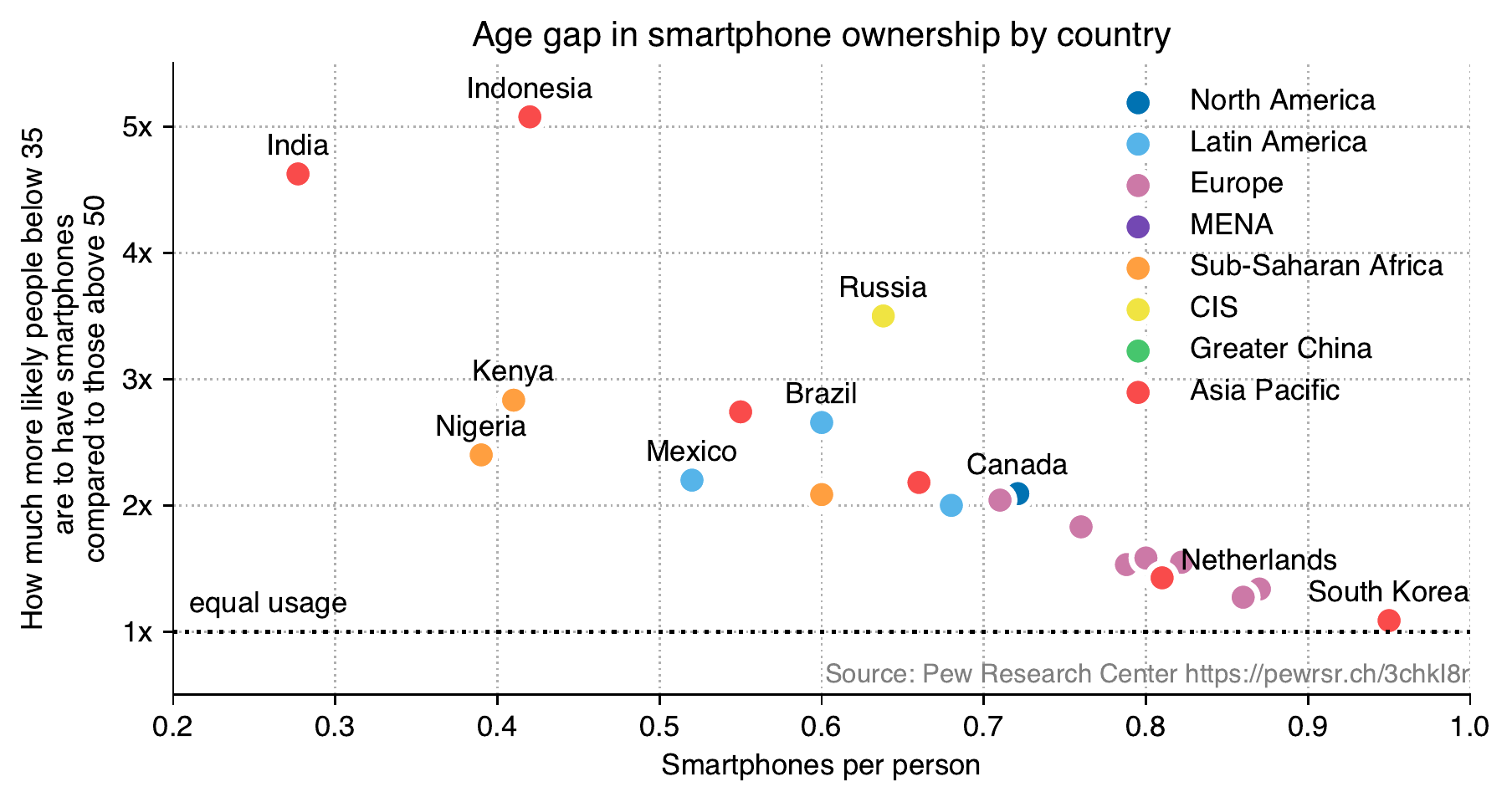}
    \caption{Countries with the lowest overall smartphone adoption also have the largest age gaps in ownership.~\cite{pew2019smartphone} In India and Indonesia, below below 35 are five times more likely to have a smartphone compared to those above 50.}
    \label{fig:age_gap}
\end{figure}

Yet, for digital contact tracing to be effective, epidemiologists estimate that we need at least a success rate of 60\% for both identifying and isolating the contacts within a few days.~\cite{ferretti2020quantifying} This fraction will differ depending on the severity of the situation (how many people are currently affected, what distancing measures are in place, etc.). It is, however, quite clear that this success rate cannot be achieved in countries where less than 60\% of the population owns a smartphone regardless of adoption levels and efficiency of the app itself. As visualised in the map, these countries are not evenly distributed around the world - instead, they include most of Latin America, the Middle East, Sub-Saharan Africa, as well as many countries in the Asia/Pacific region.

That a digital divide including uneven access to smartphones and mobile internet remains a global problem is clear. What tends to be overlooked is how these discrepancies also exist within countries. For instance, the elderly are one of the highest risk groups from a medical perspective. At the same time, they are also considerably less likely to have a smartphone compared to their young compatriots. Surveys show that in Europe general smartphone adoption is above 75\%.~\cite{pew2019smartphone} Yet, as we show in Figure~\ref{fig:age_gap}, people aged 18-35 are 1.5 times more likely to own a smartphone than those 50 years or older. This difference is especially high in countries with low general smartphone ownership. In India and Indonesia, where  penetration levels are generally low, young people are as much as five times more likely to own a smartphone than those over 50.

\begin{figure}[tp]
    \centering
    \includegraphics[width=0.75\linewidth]{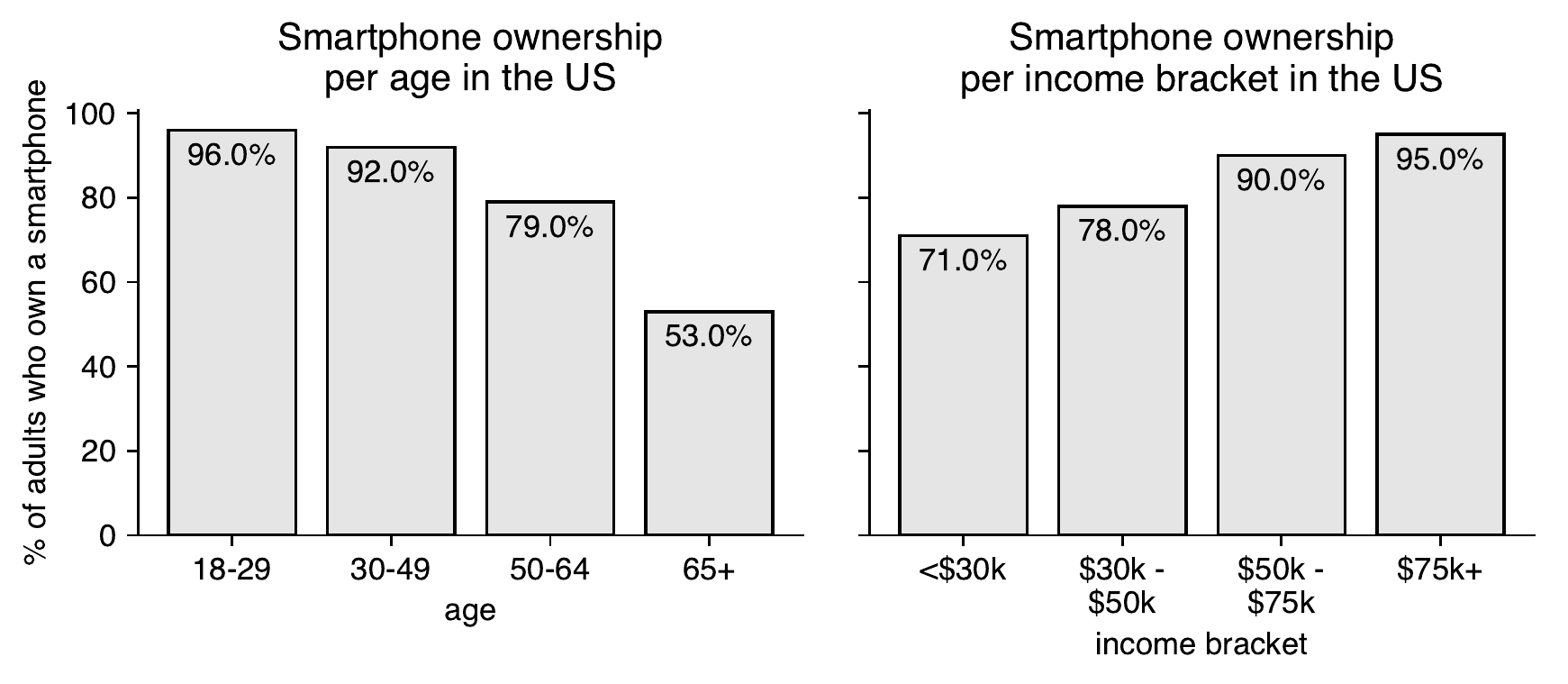}
    \caption{Smartphone adoption in the US is among the highest in the world, but those over 65+ are still much less likely to own a smartphone than those below 35.}
    \label{fig:us_ages}
\end{figure}

Even in the US where overall smartphone ownership levels  are high, we observe crucial differences between demographics. According to a Pew survey from 2019, only 53\% of Americans 65 years and older own a smartphone,~\cite{pew2019mobile} see Figure~\ref{fig:us_ages}. The number is as high as 96\% for adults between 18-29 years. That leaves 24 million people at the highest risk of dying from COVID-19 outside any smartphone based tracing system. The same Pew survey also shows that income is a decisive factor for owning a smartphone: less than three quarters of adults with a household income below \$50,000 own smartphones compared to 95\% of the those earning above \$75,000, see Figure~\ref{fig:us_ages}. It further  reports lower penetration rates among rural residents (71\% vs 83\% for (sub)urban). Finally, the Pew Research survey reports only small differences in smartphone ownership in the US among races and genders.

Obtaining detailed statistics for smartphone ownership is generally challenging and even more difficult for low and middle income countries. Because of the scarcity of such data we cannot draw as comprehensive a picture as for the US. What we can do is provide a few results for other countries to compare. For instance, in the country of Georgia (South Caucasus) the gap in smartphone ownership is strongly correlated with wealth. A recent survey observed that 95\% of the most wealthy households have at least one smartphone, while less than 40\% of the poorest households have access to one.~\cite{georgia2019} With an average household size of 3.4, smartphone adoption numbers are likely to be worse.

\begin{figure}[tb]
    \centering
    \includegraphics[width=0.75\linewidth]{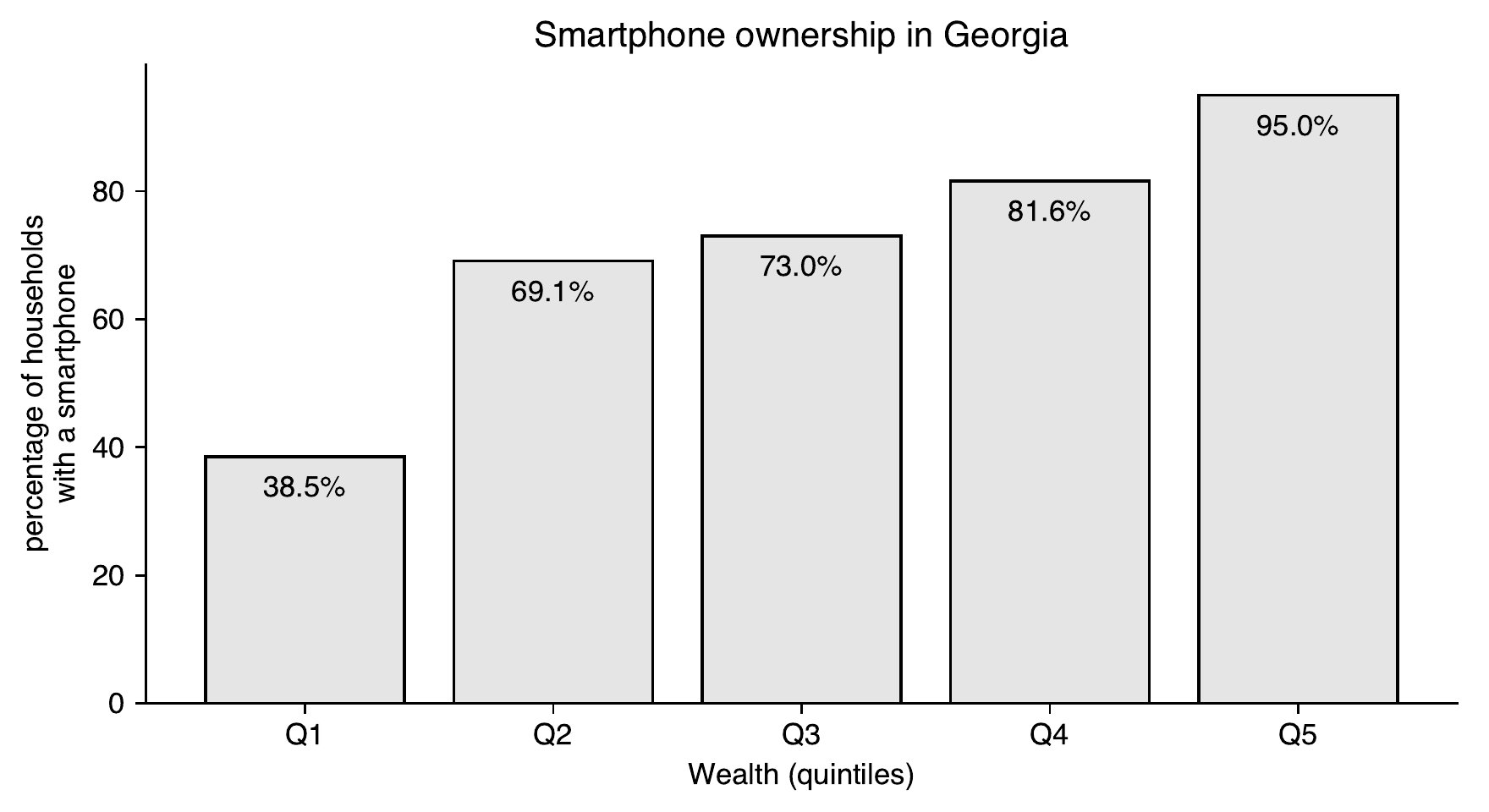}
    \caption{In the country of Georgia less than 40\% of the poorest households have even one smartphone compared to 95\% of the richest households.}
    \label{fig:georgia}
\end{figure}

Data from other countries illustrate the gender gap in access to technology. For example, in Iraq the vast majority of households own a mobile phone (this number includes both smart and feature phones), regardless of wealth and whether they live in an urban or a rural area.~\cite{iraq2019} However, when we ask the same question of women, an entirely different image emerges. Less than half of women in the poorest households and those living in rural areas own a mobile phone, let alone a smartphone. 

\begin{figure}[tb]
    \centering
    \includegraphics[width=0.75\linewidth]{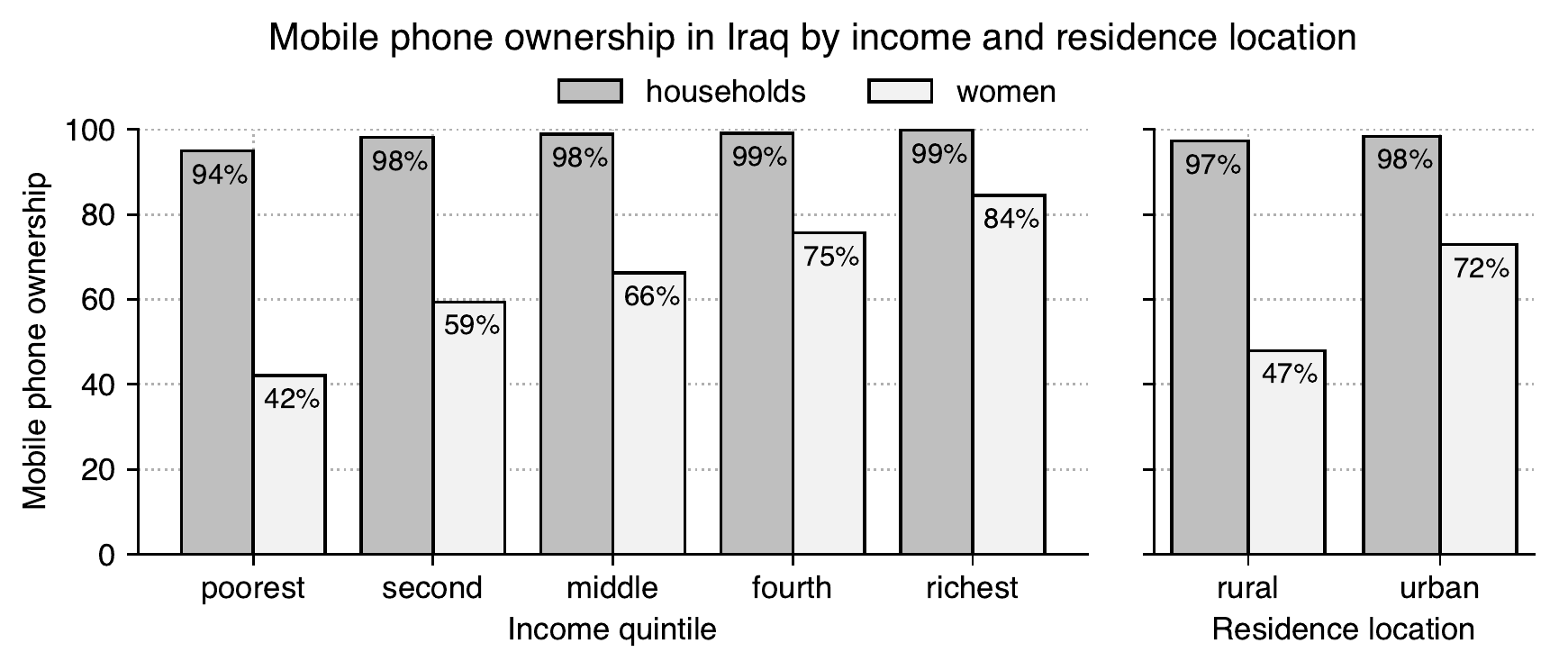}
    \caption{In Iraq the vast majority of households have a mobile phone, regardless of wealth or location. However, women, especially those in the poorest and rural households are much less likely to have a phone on their own.}
    \label{fig:iraq}
\end{figure}

A Pew Research survey from 2018 showed that while the gender gap in access to smartphones was in single digits in most of the 27 investigated countries, in India men were more than twice as likely to own one than women (34\% vs 15\%).~\cite{pew2019smartphone} Rolling out contact tracing apps without taking gender, age, wealth, ability, race, and class aspects of the digital divide  into account can thus be disastrous and exacerbate problems for the most vulnerable. 

\subsection*{A smartphone per person?}
When we think of using smartphones for contact tracing, we necessarily assume that the way people treat and use their smartphones is consistent across contexts and cultures - that smartphones belong to individuals, who constantly carry them on their bodies. As anthropologist Susan L. Erikson describes in ``Cell Phones $\ne$ Self and Other Problems with Big Data Detection and Containment during Epidemics'' this is not necessarily the case everywhere:
\begin{quote}
    \textit{Many Sierra Leoneans possess cell phones, but this possession is often temporary, even fleeting. Cell phones are loaned, traded, and passed around among family and friends, like clothes, books, and bicycles. A single phone can be shared by an extended family or, in rural areas, a neighborhood or a village. (...) Outside of larger cities, fewer Sierra Leoneans have cell phones. But in Freetown and Kenema, the capital city of the Eastern Province, I met many people who had more than one.}~\cite{erikson2018cell}
\end{quote}

The consequences of this are two-fold. First, the data on penetration rates of smartphones and mobile internet users are potentially overly optimistic: 20\% penetration does not mean two out of ten people have one smartphone each, but it might mean that 1 out of 10 has two devices. Second, even the smartphones might not be relevant  for contact tracing efforts as they do not correspond to one person but are instead shared among multiple individuals: 20\% penetration does not mean that two out of then people have one  smartphone each and the others have none. Instead, it might mean that a group of ten people share two devices and use them each at different times.

\subsection*{Internet infrastructure}
In general, smartphone users tend to have access to the Internet.~\cite{itu2019}  But the nation-wide statistics of Internet penetration might not accurately reflect the reality of ease and dependability of access. One of the limiting factors are large-scale power outages that disproportionately affect low income  countries. According to a World Bank study, countries in South Asia experience 25.5 power outages per month on average, compared to 0.3 per month in Europe.~\cite{worldbank2019}

However, power outages are not the only reason for temporary lapses in connectivity. A human rights advocacy group Access Now reports on growing numbers of deliberate Internet shutdowns across the world which predominantly affect vulnerable communities. Their \#KeepItOn campaign documented 213 shutdowns in 33 countries in 2019.~\cite{keepiton2020} Some of the recorded shutdowns only last from a few hours for example when the UK law enforcement blocked the Internet access during a climate protest. Others last up to a few months like the case of Indian government's complete blackout of Jammu and Kashmir for 175 days, or Myanmar's shutdown of the Rakhine and Rohingya communities that started in June 2019 and continues to this day.~\cite{myanmar2020}
In his recent Human Rights Council submission ``Disease pandemics and the freedom of opinion and expression'' the Special Rapporteur on the promotion and protection of the right to freedom of opinion and expression, David Kaye, stressed:

\begin{quote}
    \textit{Governments increasingly resort to shutting down the Internet, often for illegitimate purposes but in all cases having a disproportionate impact on the population. Network shutdowns invariably fail to meet the standard of necessity.}~\cite{kaye2020}
\end{quote}

None of the reasons given to justify the ongoing shutdowns are related to the spread of COVID: the authorities do so to strategically curb protests, limit the free flow of information, and silence political speech.~\cite{keepiton2020,netizen2019, indonesia2019} However, they are already hindering the pandemic response among the impacted communities for example in Kashmir,~\cite{misgar2020} and Bangladesh.~\cite{hrw2019,phr2020} The negative impact of internet shutdowns on public health response also shows us that the internet has taken up a role similar to public infrastructure. Hence, cutting Internet access does not only harm free speech and commercial ventures but harms fundamental rights such as the right to health.~\cite{UNICESCR}

\section*{Assumption 2: Bluetooth works well enough for measuring proximity of epidemiological importance}
Despite the disparities in access, some countries are already employing apps for contact tracing using  Bluetooth Low Energy. While we agree that this is the best technology we have, it is not clear whether it is good enough. Let's take a closer look.

When contact tracing is enabled on a phone, it will send out a Bluetooth signal every few seconds. Other phones around will be able to detect this signal, and measure its strength. In theory a strong signal means that the devices are near each other; the further they are, the weaker the signal.

Assuming ideal conditions with available smartphones, Bluetooth signals transmitted at full power can be picked up by other smartphones over up to 100 meters (330 ft). This is much further than the distancing advisory of 2 meters (6ft). Current app proposals rely on a combination of (1) lowering the transmission power to prevent such long distance reception and further (2) relying on Bluetooth signal strength measurement to identify \textit{short-distance} proximity events - they assume a strong signal means a short distance while a weak signal means two people were further away from each other. Unfortunately for contact tracing, this is not really the case. For example, the human body weakens  the Bluetooth signal quite well, which means that the app's distance estimate between two people talking to each other will be vastly different depending on whether their phones are in their front or back pockets. At the same time, Bluetooth signals can go through walls so your neighbors (both next door and up/down-stairs) are likely to be logged as a close contact. In fact, a few years ago we tested treating Bluetooth received signal strength' as a proxy for physical distance in ideal conditions, i.e. we used the exact same model phones, we placed them in an empty room (no human body attenuation) and various distances and found that it was very difficult to tell whether two phones were 1 meter (3 ft) apart (crucial for epidemic measurements) or 3 meters (9 ft) apart (much less relevant for transmission, as most virus-carrying droplets will fall within 2 meters, or 6ft).~\cite{sekara2014strength}

These problems will be even more difficult to solve with different phone models and different places where people carry them. Solving them will require smartphone producers to closely collaborate with Google and Apple - in each phone model the relationship between signal strength and distance will be a bit different. The code that translates signal strength to distance will need to be further calibrated for different scenarios. This translation will be different outside vs inside, for phones carried in front pockets vs back pockets, backpacks vs purses, and so on. 

Finally, there are viral transmission pathways that would not be captured  by Bluetooth. For example a widely discussed paper describes viral transmission among people in a restaurant.~\cite{luearly} Some of the infected guests were  not sitting in close proximity to the infected person but instead droplets were transported through the air duct. Bluetooth apps would also not cover possible transmission through touching a common surface, with the caveat that we still do not know how likely this transmission channel is.

Like every technology whose purpose is to detect certain events, Bluetooth-based contact tracing will suffer from two kinds of errors: \textit{false negatives} and \textit{false positives}. 
False negatives represent the failure of the app to report a potentially dangerous contact, for example because the distance was misjudged, or even because of external factors: the battery died, or the infected person did not have a smartphone. 
False positives happen when the app reports a potential exposure to the virus even though such exposure was very unlikely, for example the phones detected each other through a wall.
These two kinds of errors have the potential to further drive societal inequities as they affect certain groups of people unequally.
False negatives will mean that those who do not have smartphones or choose not to install the app might find it more difficult to obtain testing and adequate healthcare.
False positives will predominantly affect those living with more people, take public transport, and work essential jobs that require them to be around others. 
If they are expected to isolate after receiving an exposure warning, it might lead to inequitable quarantining~\cite{landau2020} and undermine trust.

If Bluetooth is so inaccurate, why do academics (including authors of this work) keep using it as a proxy for human interactions also in their work on modeling the spread of diseases? In short, it is the best we have at the moment. Bluetooth is still much more accurate than GPS data and while not perfect, the signal still tells us something about physical proximity. Most of the academic work with this data revolves around simulations of spread dynamics. Rather than tracking who exactly was exposed, academics are interested in probabilistic estimates of how quickly the disease will spread, how many people will be infected, and so on. When we do focus on the risk to particular individuals, we get to run dozens of thousands of simulations to estimate the probability of each person being infected. Only when the system is deployed in the real world, will we have a chance to learn whether this probability corresponds to real infections well enough to support manual contact tracing efforts without overwhelming the system with false positives. When making such comparisons, we need to remember that currently in many places only people who have symptoms and were identified through contact tracing are tested,~\cite{markup2020} and still the majority of them turn out not to have the virus.

\section*{Assumption 3: People can and will choose to install and use contact tracing apps}

Digital rights organizations around the world stress that contact tracing technology needs to respect and protect civil liberties. In this ideal scenario, adoption of any contact tracing technology would be voluntary, based on informed consent and trust in the apps' efficacy and the institutions governing them. This trust should be built through guarantees regarding transparency (of code, of operation, but also of decision making processes and public procurement), legality, well enforced data protection standards, and accountability.~\cite{article19}

However, we have seen both democratic and non-democratic governments rushing to implement tracing apps and centralised data collection systems (see \url{https://privacyinternational.org/examples/apps-and-covid-19}). Many initially proposed tools suffer from mission creep and are designed and/or operated by private companies, lacking public oversight, a defined legal framework and adequate privacy protections.~\cite{osborne2020} 
%Google and Apple essentially rolled out public health policy as an update to their operating systems, without public debates and a democratic deliberation process. 
The result of this situation is a lack of public trust expressed in a low willingness to opt in and install the existing apps. A poll from April 2020 shows that Americans have low levels of trust in tech companies, universities, insurance companies, and public health agencies. As a result, half of those who have smartphones in the US will not install a contact tracing app.~\cite{wapo2020}

\begin{figure}[t]
    \centering
    \includegraphics[width=0.75\linewidth]{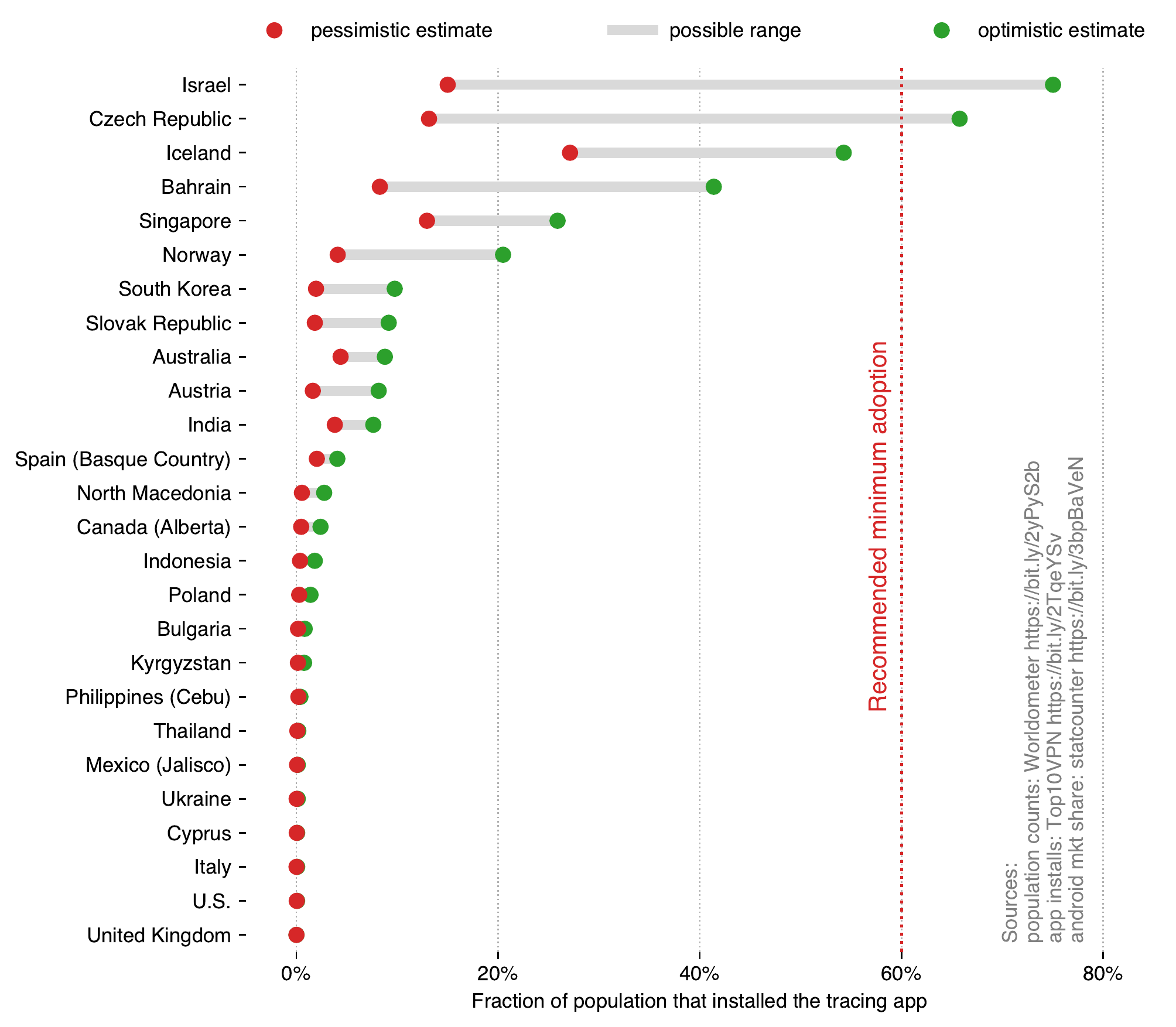}
    \caption{Despite contact tracing apps being deployed in 30 countries, the adoption levels are disappointing. Only Israel and Czech Republic might have reached the required level of 60\%, assuming the most optimistic estimate. }
    \label{fig:adoption}
\end{figure}

Experiences from around the world seem to reflect these findings. The first country to roll out an official and voluntary contact tracing app was Singapore with \texttt{TraceTogether}. Popular app uptake proved to be disappointing, despite promises of data minimisation and privacy safeguards, and the opening of the app source code. Six weeks after deployment only 1.1 million (or less than one in five) residents installed it. That is far below the minimum goal of 75\%, set by their national response team.~\cite{chong2020} Norway's experience was similar. Two weeks after the release of their contact tracing app it had 860,000 users, corresponding to only 16\% of the population.~\cite{olsen2020} Such low adoption in small populations known for high levels of trust in their governments does not inspire confidence in adoption elsewhere. Unfortunately, we do not have accurate data about active users of other tracing apps. Apple does not publish download statistics and Google only provides a rough estimate. And even with that data we still would not know how many downloads actually translate into active users. Our approximations presented in Figure~\ref{fig:adoption} show that adoption is, in fact, very low across the board (see Section Materials and Methods at the end of this article to find out how we arrived at these estimates). At the time of this analysis (mid-May 2020), we clearly see that only two countries, Israel and Czech Republic, \textit{might} have reached the 60\% threshold of installs that is commonly referred to as the minimum for tracing to work, and that is only if we make the most optimistic assumption. 

Moreover, a variety of groups across the US and Europe outright reject governmental containment efforts.
Some actively resist lockdowns and the wearing of masks.~\cite{bbc2020,fa2020} Ultra-Orthodox Jews~\cite{ettinger2016} and the Amish tend to refrain from technologies on religious grounds. It seems unlikely that these groups will install a contact tracing tool voluntarily, thus creating pockets of population with significantly lower adoption than the general public. 

\subsection*{Enforcing the use of apps}
As adoption levels fail to materialise, some governments started to mandate apps by law or make them a precondition to full participation in public life. In the latter case an app might not be legally mandatory and still based on \textit{consent}, but that consent is not meaningful if the reward is too high for citizens to actually be able to say no. While international law allows for limitations to civil liberties during a public health crisis, including on the right to privacy, it also clearly defines how the government should implement such limitations to balance rights and prevent abuses.~\cite{hrc2020} It is yet unclear whether   directly or indirectly enforcing proposed tools would actually withstand legal tests.~\cite{reventlow2020} A coercive approach is also problematic if it ignores the fact that many people might not have internet access or smartphones as shown in Assumption 1. 

Notwithstanding these concerns, the Indian government made their contact tracing app mandatory for workers across the country and for everybody in select cities (resulting in 90 million downloads corresponding to less than 7\% of the population).~\cite{phartiyal2020}
Poland's quarantine app is obligatory despite its privacy flaws.~\cite{nowosielska2020}
Calls for mandatory app use are gaining traction in Singapore~\cite{tham2020} and Italy.
South Korea requires foreigners to install a tracking app when entering the country.~\cite{smith2020} 
Kazakhstan,~\cite{zaik2020} Russia,~\cite{nesterova2020} and the Gulf States~\cite{frank2020} require installing tracking tools as a prerequisite for people to leave the house during quarantine. 
To understand what mandating  really means for people's lives, it is worth looking at enforcement actors and practices. 
India~\cite{dixit2020b} and Poland~\cite{govpl2020} task the police with enforcing tools through fines or even jail time. 
In Russia, the enforcement is built into the quarantine app: failure to comply with the instructions given by the app automatically results in a fine, even if it's caused by a bug in the app itself.~\cite{marohovskaja2020} We also have major concerns about how enforcement will look like for people who do not have phones or the Internet to comply, or for people whose economic situation forces them to keep leaving the house. Looking at patterns of quarantine enforcement to date, we can expect that extended police powers will also aggravate already dire patterns of over-policing, racism,~\cite{southall2020} and police brutality.~\cite{olewe2020}

\subsection*{The future of the surveillance infrastructure}
Finally, we are also seeing how the crisis is used as a cover to further extend surveillance powers and introduce unrelated infrastructure often targeting the already over-surveilled. 
Reportedly, China intensified hacking campaigns against the Uighur community~\cite{greenberg2020} and started installing cameras outside and sometimes inside people's homes.~\cite{gan2020} 
Israel requires Palestinians who want to ensure the validity of their permit to stay in Israel to install an app giving the military access to their phone data.~\cite{hasson2020}  
Moscow introduced drones to police quarantine orders~\cite{vedomosti2020} and Paris has adopted AI tools to assess mask adoption on public transport.~\cite{vincent2020}
The US Senate has passed a law allowing the FBI to access the browsing history of all Americans without a warrant~\cite{rose2020} in times where we all have to rely on the internet to get on with our lives. 
And finally, the police in South Korean and Hong Kong track people's movements with wristbands, a tool often considered as a potential alternative to apps including in India and Bulgaria.~\cite{bbc2020wrist}
The list is growing by the day.

These examples of deploying over-reaching surveillance tech under the guise of COVID-19 response are already alarming. More subtle tech interventions like privacy-preserving Bluetooth based contact tracing might seem benign by comparison but we will only understand their full impact in the years to come. We need to stay vigilant regarding re-purposing of tools and data.  Sean McDonald points out that:
\begin{quote}
    \textit{The most political dimension of contact-tracing apps isn't what they're being used for right now, it's the role they are likely to play in attempts to reopen societies.}\cite{mcdonald2020}
\end{quote}

Recent history has shown that it is extremely difficult to roll back overextended government powers once they have been institutionalized and normalised. In the wake of the 9/11 attacks, the U.S. introduced the Patriot Act to drastically extend their surveillance apparatus with the goal of curbing terrorism. We do not know how effective it was in its stated objective, but, through the Snowden leaks, we do know that it has been widely abused to encroach on the rights of people worldwide.~\cite{bbc2014} 19 years after the attacks, the Patriot Act has still not been repealed, setting a dangerous precedence. We thus need to carefully analyse what role contact tracing tools can play in a given power equilibrium and what is their potential for retooling once the pandemic is contained.

\section*{Assumption 4: Exposed individuals can adapt their behavior accordingly}
Experts claim between as many as 20 million tests \textit{per day} will be necessary to safely reopen the US,~\cite{allen2020roadmap} (corresponding to 60 tests per thousand residents per day) but the current (as of May, 20 2020) \textit{total for the entire epidemic} is only at 12,647,791\cite{dashboard2019} (39 per thousand total). Why do we need so many?
Without the tests we will not know whose contacts to trace. Moreover, if we cannot test these contacts, we will need to isolate them. At this stage of the epidemic, with a sizable infected population, isolating all contacts would not be much different than a complete lockdown. As the current restrictions are lifted each carrier might cross paths (and spend some time) with dozens of susceptible people daily but only a fraction of them will actually become infected - we need to be able to test all of them, but isolate only those that do contract the virus. 

\begin{figure}[t]
    \centering
    \includegraphics[width=0.75\linewidth]{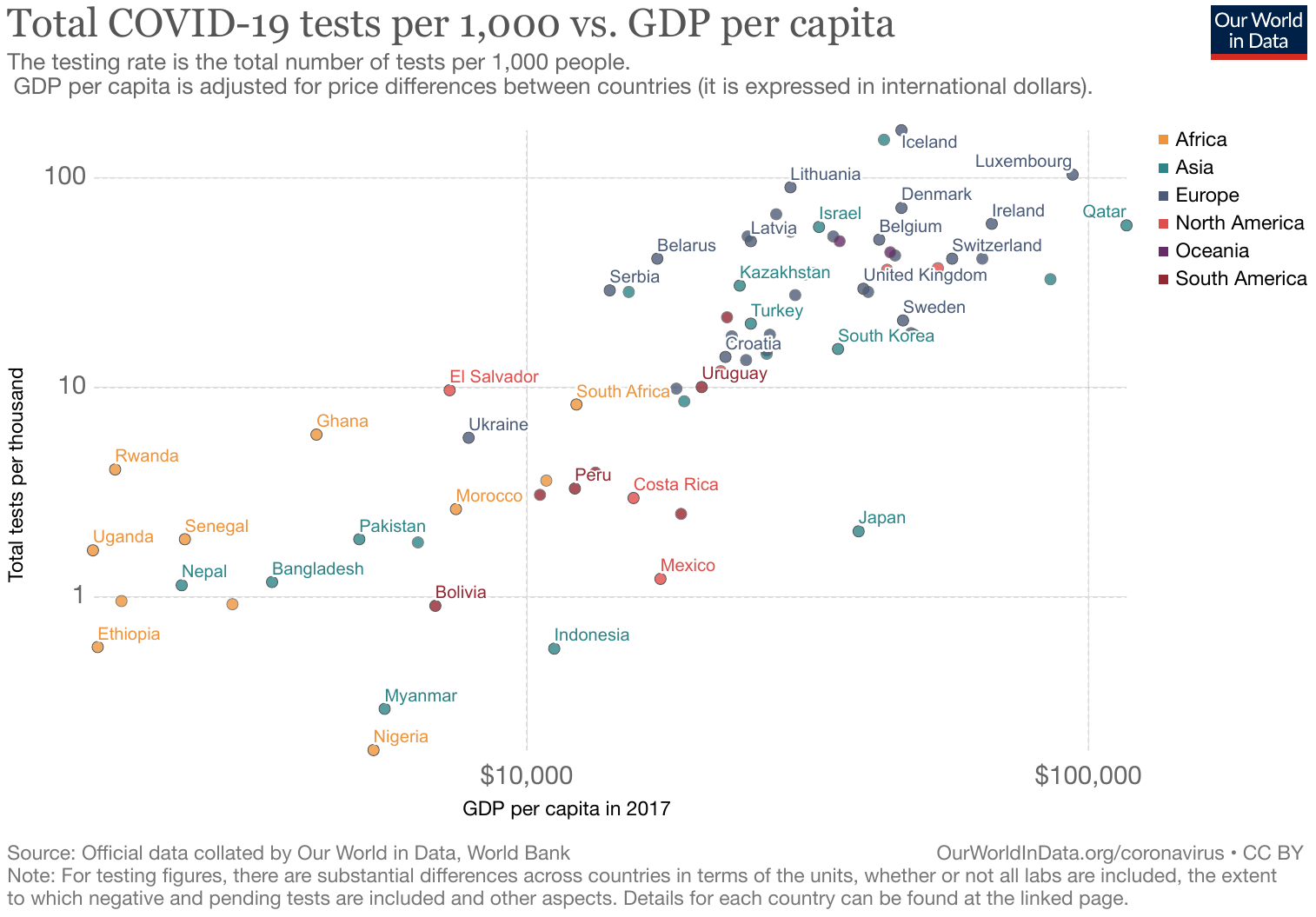}
    \caption{Most countries struggle with performing adequate number of tests, but lower income countries are especially affected by the shortages.}
    \label{fig:testing}
\end{figure}

Unfortunately, the shortage of testing, lack of smartphones and internet access, as well as share of informal employment all come together to disproportionately impact the lower income countries, see Figure~\ref{fig:testing}. The lack of tests and limited access to those available drives the health divide rooted in social status and economic differences even further. There are three factors at play: first, the economic feasibility of ceasing to work or working from home while isolating; second, the efficacy of staying at home in order to slow down the spread; and third the ability to isolate at all.

Wealthier people will isolate at their homes (or summer houses~\cite{hoffower2020}), but not everybody can afford to not work, or work from home. WHO guidelines state that individuals expected to stay at home
\begin{quote}
    \textit{...need to be provided with health care; financial, social and psychosocial support; and basic needs, including food, water, and other essentials. The needs of vulnerable populations should be prioritized.}~\cite{world2020considerations}
\end{quote} 

Unfortunately, this is not the reality for a large fraction of the world's population. The latest brief of the International Labour Organisation reminds us that the 2 billion informal workers worldwide are faced
\begin{quote}
    \textit{..with an almost unsolvable dilemma: to die from hunger or from the virus.}\cite{ilo2020}
\end{quote}
This holds also true in richer countries. In the US, only two months into non-essential business closures, food insecurity has reached an unprecedented level. According to the COVID Impact Survey and the Survey of Mothers with Young Children,~\cite{bauer2020} two in five households run by mothers with children up to 12 years old could not afford enough food, more than a two-fold increase compared to the same time of year in 2018.

\begin{figure}[tb]
    \centering
    \includegraphics[width=0.75\linewidth]{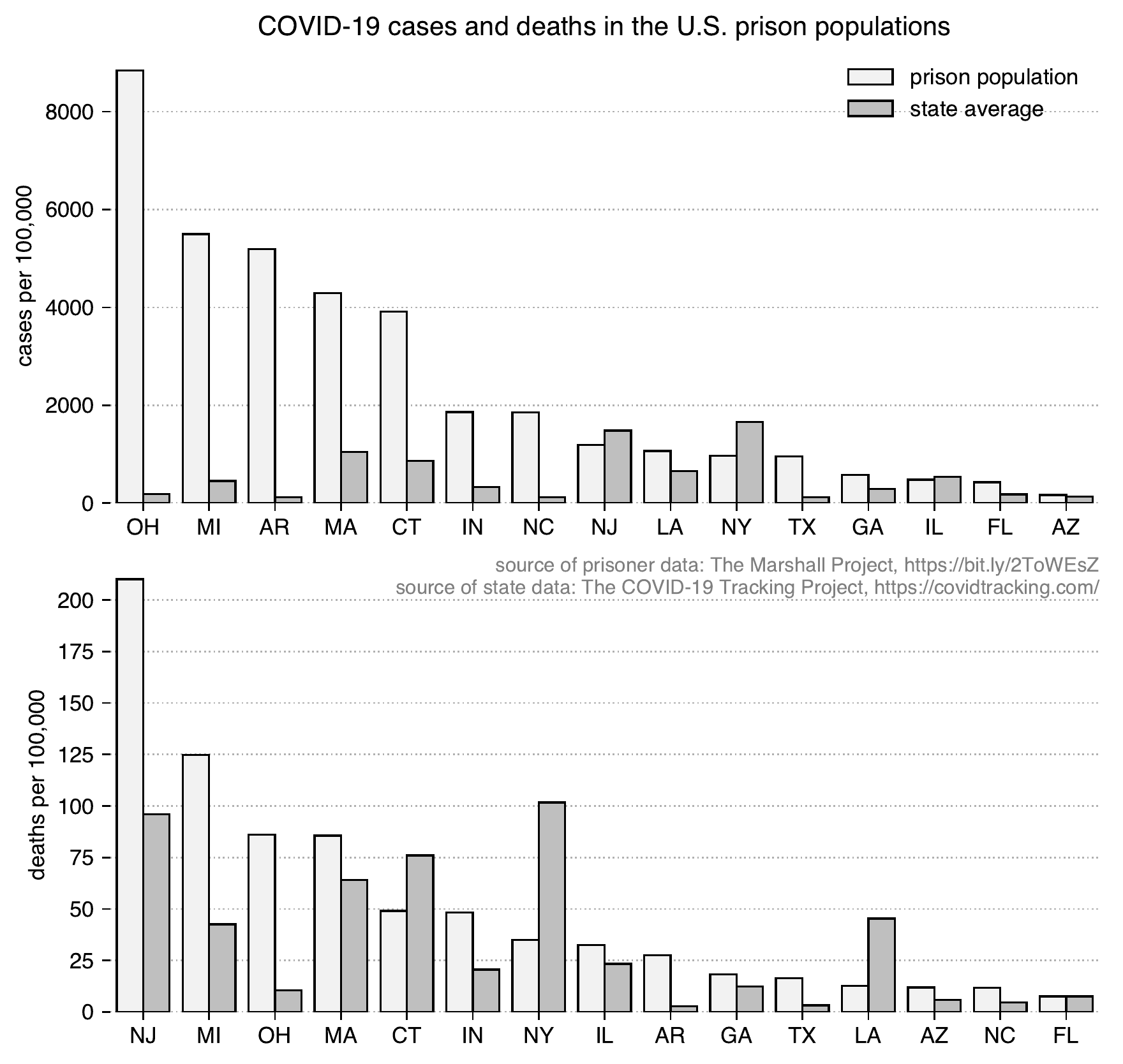}
    \caption{Imprisoned populations cannot practice social distancing and might suffer from worse access to health care. As a result, the rates of infection and deaths are much higher than average for the states where the prisons are located.}
    \label{fig:prisons}
\end{figure}

For many across the world staying at home does not only endanger their food supply, but might also not contribute much to slowing the virus' spread. Transmission between people housed together is now a major driver of the pandemic.~\cite{cevik2020}
Those with lower incomes are more likely to share smaller living spaces with more people, often with insufficient sanitary conditions. One in seven people on Earth live in a slum,~\cite{unstats2019} where the population density can be as much as ten times higher than in the surrounding city.~\cite{bird2017life} Additionally, More than half of the world's population does not have access to sanitation.~\cite{who2019} Without being able to socially distance and practice hygiene, just staying at home will not stop the virus - sadly many of these areas are already suffering COVID outbreaks.  

Finally, some communities are not free to change their behavior and seek help in response to the virus. Roma settlements in Bulgaria and Slovakia were cordoned off without providing adequate health and social aid.~\cite{walker2020} The virus also ravages migrant workers' dorms~\cite{hadavas2020} and refugee camps.~\cite{subbaraman2020distancing} While we lack sufficient data in these contexts (in part because of the lack of testing), we do know what happens once the virus reaches another kind of population that cannot alter its behavior accordingly: those held in prisons in the US. Once the virus makes it through the gates, it spreads much more rapidly than in the rest of the population, leading to enormous differences in infections and deaths. For example, as we show in Figure~\ref{fig:prisons} based on the data collected by The Marshall Project,~\cite{marshall2020} in Ohio prisons we have seen 48 times more cases and 8 times more deaths per 100,000 than the state average. These numbers are similar for Arkansas: we see 44 times more cases and 10 times more deaths compared to the state average.

All of the factors we describe above are inherently connected. Economic hardship tends to overlap with other vulnerabilities and aggravates them. Living in conflict areas or fragile states; living with displacement or an undetermined migration status; even living in richer countries but lower on the socio-economic ladder or struggling with homelessness; all of these situations often also mean sharing narrow spaces with insufficient sanitary conditions, and limited agency to adapt individual behavior to protect oneself, the family, and the community.

Without public health responses pro-actively addressing these needs of vulnerable communities, many of those affected by the virus will neither be able to access testing nor isolate. As a result, the virus is likely to continue spreading even if contact tracing apps will reach high levels of adoption in the general population. No app will help when those affected cannot respond accordingly to exposure notifications. All that tracing tools can do in these contexts is collect data, increase surveillance, and add an additional mental burden to an already incredibly challenging situation.

\section*{Conclusions}
In this commentary we took a step back from the discussion about the intricacies of the contact tracing technology and focused on challenging the assumptions that are too often taken for granted. 
By looking at recent data and latest reporting, analysis and commentary from around the world we can confidently say: 
\begin{itemize}
\item Estimates show that over half of the global population do not have modern smartphones, unhindered access to the Internet, and incompatible user patterns. 
\item Inevitable errors in Bluetooth sensing will affect populations differently: those without smartphones might have a harder time accessing appropriate health care if lines for testing are filled with app users falsely alarmed about potential exposure.
\item Growing evidence shows that individuals rarely choose to install tracing apps unless coerced in some way, raising serious concerns in regards to civil liberties and the long term impact of these surveillance systems.
\item The majority of the world's population cannot access tests and many cannot adapt their behavior accordingly if they were notified about exposure due to socioeconomic factors and other vulnerabilities. 
\end{itemize}

It is abundantly clear that no technology exists in a political, social, and economic vacuum and that the current epidemic disproportionately harms the already vulnerable communities. Basing solutions to the problem on the technology that is available mostly to the younger, better educated, richer and/or living in the rich countries might help them, might also bring unintended and potentially harmful consequences, discriminatory enforcement practices  and further exacerbate existing inequalities.

Reliance on apps is, however, not just a problem of access to technology. So far, we have seen these apps designed and deployed without conducting human right impact assessments or allowing public discussion and scrutiny. We must also consider how the national deployments, mandating, and enforcement will affect those that do install the apps, or wear wristbands. Seda G{\"u}rses emphasises the existence of two distinct classes in the society. The \textit{delivery class} delivers the goods and health care to the \textit{receiver class}.~\cite{gurses2020} Individuals in the latter can afford to stay at home and will neither be at a heightened risk of the virus, nor exposed to the inherent privacy risks of running surveillance apps on their phones. The delivery class, on the other hand, will be expected, or even forced, to install the apps and carry the biggest burden both of the pandemic but also of privacy loss, as well as intensified scrutiny and surveillance from law enforcement.  The voices of the delivery class are strikingly absent in the debates on how to design public health responses in general and technology in particular.

Smartphone contact tracing promises to  \textit{support} manual tracers, not  \textit{replace} them. In fact, the very people who are developing these apps are urging policymakers to strengthen their \textit{offline} response and hire more tracers.~\cite{salathe2020} But is this really the situation we are working towards? 
Some countries do: Georgia,~\cite{georgia2020} Ghana, and Senegal,~\cite{ghana2020} as well as India's state of Kerala~\cite{diplomat2020} are successfully managing to slow down the virus with considerate and rapid public health responses and without enforcing new, untested technology.
In the US currently only one state, North Dakota, has an adequate number of tracing personnel,~\cite{duffin2020} and only a few more are planning to meet the demand. With millions of people around the world losing their jobs in the wake of the pandemic, \textit{supporting} the manual tracing efforts could provide economic relief to individuals and help contain the spread beyond the digital divide, before experimenting with unproven and potentially harmful tech fixes.

\section*{Materials and Methods}
\noindent The data for Figure~\ref{fig:gsma_map}  as well as allocation of countries to market regions, come from the GSMA The Mobile Economy report accessible at \url{https://bit.ly/3bmYrol}. The particular numbers are transcribed from Figure 6 on page 14. They refer to ``mobile broadband'' subscribers rather than ``smartphone ownership'', so they involve all those that have a SIM card with mobile data capability even if they do not have a smartphone. 
\vspace{.5cm}

\noindent There are two sources of smartphone penetration data in this Figure~\ref{fig:ownership}: a Pew Research Center survey from 2019 (\url{https://pewrsr.ch/3chkI8r}) and a Newzoo study from 2019 (\url{https://bit.ly/2WCnUF3}). Whenever a country appeared in both surveys, we took the more optimistic (higher) estimate of penetration. The GDP estimates were taken from 2018 (the year when both penetration studies were performed) from The World Bank \url{https://bit.ly/2LjyRX0}. The population estimates from 2018 were from the same Newzoo study for countries it covered and from the United Nations \url{https://bit.ly/3dGgK9S} for the other countries.
\vspace{.5cm}

\noindent All the data in Figure~\ref{fig:age_gap} comes from the Pew Research Center study (\url{https://pewrsr.ch/2LotKVC}) from page 13. Y-axis is the result of dividing penetration among the 18-34 year olds by the penetration among the 50+ year old. X-axis is the country's total penetration.
\vspace{.5cm}

\noindent To arrive at the results presented in Figure~\ref{fig:adoption} we took the following steps. First, we used the install estimates from Google Play store as collected and reported by Top10VPN at \url{https://bit.ly/2TqeYSv}. These estimates come in ranges 5-10, 11-50, 51-100, 101-500, and so on. The Top10VPN data also reports if each app has an iOS version. If it does, we use Statcounter to obtain the estimate of Android market share in that country. We assume, that the same fraction of iOS users installs the app as Android users. We then report the pessimistic estimate as the lower bound reported by Google Play Store plus the corresponding number of installs from iOS users, and the optimistic estimate as upper bound plus the corresponding number of iOS installs, both divided by the country's population.

\section*{Acknowledgements}
The authors would like to thank all those who volunteered their time to help us make this article better: NaLette Brodnax, Dan Calacci, Manuel Garcia-Herranz, Enys Mones, Seda Gürses, Alan Mislove, Alissa Sadler, Rachel Ross, Jonathan Andrew, and others.

\section*{Authors}
\noindent\textbf{Piotr Sapiezynski} is a Research Scientist at Northeastern University, working on societal impacts of algorithmic systems. His PhD work focused on analysing human mobility and physical proximity patterns through the lens of data collected using smartphones. He has published on the topic in top conferences and journals.
\vspace{.5cm}

\noindent\textbf{Johanna Pruessing} has a Master's in anthropology and political science. She has lived and worked in Russia, Armenia, Azerbaijan, Turkey, the UK, and Poland focusing on human rights issues before joining the Open Society Foundations in Berlin as a program specialist concentrating on former Soviet countries. Thematically she works at the intersection of human rights, activism, and technology with a sweet spot for fighting digital authoritarianism.
\vspace{.5cm}

\noindent\textbf{Vedran Sekara} is joining the IT University in Copenhagen as an Assistant Professor and is a research consultant at UNICEF. His work focuses on data representativeness, building equitable algorithms, and identifying and fixing bias in common Data Science, Machine Learning, and Artificial Intelligence like systems. He has a PhD in applied mathematics and computer science which focuses on using digital traces collected from mobile phones to understand social relations. 

\bibliography{sample}

\begin{thebibliography}{10}
\urlstyle{rm}
\expandafter\ifx\csname url\endcsname\relax
  \def\url#1{\texttt{#1}}\fi
\expandafter\ifx\csname urlprefix\endcsname\relax\def\urlprefix{URL }\fi
\expandafter\ifx\csname doiprefix\endcsname\relax\def\doiprefix{DOI: }\fi
\providecommand{\bibinfo}[2]{#2}
\providecommand{\eprint}[2][]{\url{#2}}

\bibitem{vogelstein2020}
\bibinfo{author}{Vogelstein, F.}
\newblock \bibinfo{journal}{\bibinfo{title}{{Health Officials Say 'No Thanks'
  to Contact-Tracing Tech}}}.
\newblock {\emph{\JournalTitle{WIRED}}}  (\bibinfo{year}{2020}).

\bibitem{who2020}
\bibinfo{author}{{World Health Organization}}.
\newblock \bibinfo{journal}{\bibinfo{title}{Situation report - 73}}.
\newblock {\emph{\JournalTitle{{Coronavirus disease 2019 (COVID-19)}}}}
  (\bibinfo{year}{2020}).

\bibitem{goodin2020}
\bibinfo{author}{Goodin, D.}
\newblock \bibinfo{journal}{\bibinfo{title}{{Apple and Google detail bold and
  ambitious plan to track COVID-19 at scale}}}.
\newblock {\emph{\JournalTitle{{Ars Technica}}}}  (\bibinfo{year}{2020}).

\bibitem{case2020}
\bibinfo{author}{Case, N.}
\newblock \bibinfo{title}{Protecting lives \& liberty,
  \url{https://ncase.me/contact-tracing/}} (\bibinfo{year}{2020}).

\bibitem{cuomo2020}
\bibinfo{author}{{Andrew Cuomo (@NYGovCuomo)}}.
\newblock \bibinfo{title}{{[Tweet],
  {\url{https://twitter.com/NYGovCuomo/status/1245021319646904320} }}}
  (\bibinfo{year}{2020}).

\bibitem{diva2020}
\bibinfo{author}{{90s Kinda Girl (@divafeminist)}}.
\newblock \bibinfo{title}{{[Tweet],
  {\url{https://twitter.com/divafeminist/status/1256927274936553473} }}}
  (\bibinfo{year}{2020}).

\bibitem{odea2020}
\bibinfo{author}{O'Dea, S.}
\newblock \bibinfo{journal}{\bibinfo{title}{Smartphone users worldwide
  2016-2021}}.
\newblock {\emph{\JournalTitle{Statista}}}  (\bibinfo{year}{2020}).

\bibitem{bradshaw2020}
\bibinfo{author}{Bradshaw, T.}
\newblock \bibinfo{journal}{\bibinfo{title}{{2 billion phones cannot use Google
  and Apple contact-tracing tech}}}.
\newblock {\emph{\JournalTitle{{Ars Technica}}}}  (\bibinfo{year}{2020}).

\bibitem{itu2020}
\bibinfo{author}{{ITU}}.
\newblock \bibinfo{title}{{Facts and figures 2019 - Measuring digital
  development}}.
\newblock
  \bibinfo{howpublished}{\url{https://itu.foleon.com/itu/measuring-digital-development/home/}}
  (\bibinfo{year}{2020}).

\bibitem{pew2019smartphone}
\bibinfo{author}{Silver, L.}
\newblock \bibinfo{journal}{\bibinfo{title}{{Smartphone Ownership Is Growing
  Rapidly Around the World, but Not Always Equally}}}.
\newblock {\emph{\JournalTitle{Pew Research Center}}}  (\bibinfo{year}{2019}).

\bibitem{ferretti2020quantifying}
\bibinfo{author}{Ferretti, L.} \emph{et~al.}
\newblock \bibinfo{journal}{\bibinfo{title}{Quantifying sars-cov-2 transmission
  suggests epidemic control with digital contact tracing}}.
\newblock {\emph{\JournalTitle{Science}}} \textbf{\bibinfo{volume}{368}}
  (\bibinfo{year}{2020}).

\bibitem{pew2019mobile}
\bibinfo{author}{Anderson, M.}
\newblock \bibinfo{journal}{\bibinfo{title}{{Mobile Technology and Home
  Broadband 2019}}}.
\newblock {\emph{\JournalTitle{Pew Research Center}}}  (\bibinfo{year}{2019}).

\bibitem{georgia2019}
\bibinfo{author}{{UNICEF, National Statistics Office of Georgia}}.
\newblock \bibinfo{title}{{ Georgia Multiple Indicator Cluster Survey 2018 -
  Survey Findings Report}}.
\newblock \bibinfo{type}{Tech. Rep.}, \bibinfo{institution}{{UNICEF}}
  (\bibinfo{year}{2019}).

\bibitem{iraq2019}
\bibinfo{author}{{UNICEF, Central Statistical Office, and Kurdish Region
  Statistical Office}}.
\newblock \bibinfo{title}{{Iraq Georgia Multiple Indicator Cluster Survey 2018
  - Survey Findings Report}}.
\newblock \bibinfo{type}{Tech. Rep.}, \bibinfo{institution}{{UNICEF}}
  (\bibinfo{year}{2019}).

\bibitem{erikson2018cell}
\bibinfo{author}{Erikson, S.~L.}
\newblock \bibinfo{journal}{\bibinfo{title}{Cell phones$\ne$ self and other
  problems with big data detection and containment during epidemics}}.
\newblock {\emph{\JournalTitle{Medical anthropology quarterly}}}
  \textbf{\bibinfo{volume}{32}}, \bibinfo{pages}{315--339}
  (\bibinfo{year}{2018}).

\bibitem{itu2019}
\bibinfo{author}{{ITU}}.
\newblock \bibinfo{title}{{Percentage of Individuals using the Internet}}.
\newblock
  \bibinfo{howpublished}{\url{https://www.itu.int/en/ITU-D/Statistics/Documents/statistics/2019/Individuals_Internet_2000-2018_Dec2019.xls}}
  (\bibinfo{year}{2019}).

\bibitem{worldbank2019}
\bibinfo{author}{{The World Bank}}.
\newblock \bibinfo{title}{{Power outages in firms in a typical month
  (number)}}.
\newblock
  \bibinfo{howpublished}{\url{https://data.worldbank.org/indicator/IC.ELC.OUTG}}
  (\bibinfo{year}{2019}).

\bibitem{keepiton2020}
\bibinfo{author}{{\#KeepItOn Campaign}}.
\newblock \bibinfo{journal}{\bibinfo{title}{Targeted, cut off, and left in the
  dark}}.
\newblock {\emph{\JournalTitle{AccessNow}}}  (\bibinfo{year}{2020}).

\bibitem{myanmar2020}
\bibinfo{author}{Lee, Y.}, \bibinfo{author}{Jimenez-Damary, C.},
  \bibinfo{author}{Kaye, D.} \& \bibinfo{author}{de~Varennes, F.}
\newblock \bibinfo{journal}{\bibinfo{title}{{UN} experts concerned at surge in
  civilian casualties in northwest myanmar after internet shutdown}}.
\newblock {\emph{\JournalTitle{{United Nations Human Rights, Office of the High
  Commisioner}}}}  (\bibinfo{year}{2020}).

\bibitem{kaye2020}
\bibinfo{author}{Kaye, D.}
\newblock \bibinfo{title}{{Report of the Special Rapporteur on the promotion
  and protection of the right to freedom of opinion and expression,
  {A/HRC/35/22. (para 14)}}} (\bibinfo{year}{2020}).

\bibitem{netizen2019}
\bibinfo{author}{Abrougui, A.} \emph{et~al.}
\newblock \bibinfo{journal}{\bibinfo{title}{{Netizen Report: Widespread
  throttling puts social media out of reach in Kazakhstan}}}.
\newblock {\emph{\JournalTitle{GlobalVoices}}}  (\bibinfo{year}{2019}).

\bibitem{indonesia2019}
\bibinfo{author}{Carolina, J.}
\newblock \bibinfo{journal}{\bibinfo{title}{{Indonesia's post-election riots
  led to free speech violations }}}.
\newblock {\emph{\JournalTitle{{GlobalVoices - advox}}}}
  (\bibinfo{year}{2019}).

\bibitem{misgar2020}
\bibinfo{author}{Misgar, U.~L.}
\newblock \bibinfo{journal}{\bibinfo{title}{{In Kashmir, slow internet
  throttles doctors’ coronavirus response}}}.
\newblock {\emph{\JournalTitle{The New Humanitarian}}}  (\bibinfo{year}{2020}).

\bibitem{hrw2019}
\bibinfo{author}{{Human Rights Watch}}.
\newblock \bibinfo{title}{{Bangladesh: Internet Blackout on Rohingya
  Refugees}}.
\newblock \bibinfo{type}{Tech. Rep.}, \bibinfo{institution}{{Human Rights
  Watch}} (\bibinfo{year}{2019}).

\bibitem{phr2020}
\bibinfo{author}{{Kine, Phelim}}.
\newblock \bibinfo{journal}{\bibinfo{title}{{Internet Curbs on Rohingya Risk
  Wider Virus Outbreak}}}.
\newblock {\emph{\JournalTitle{{Physicians for Human Rights}}}}
  (\bibinfo{year}{2020}).

\bibitem{UNICESCR}
\bibinfo{author}{{The United Nations General Assembly}}.
\newblock \bibinfo{journal}{\bibinfo{title}{{International Covenant on
  Economic, Social, and Cultural Rights (Article 12)}}}.
\newblock {\emph{\JournalTitle{{Treaty Series}}}}
  \textbf{\bibinfo{volume}{999}} (\bibinfo{year}{1966}).

\bibitem{sekara2014strength}
\bibinfo{author}{Sekara, V.} \& \bibinfo{author}{Lehmann, S.}
\newblock \bibinfo{journal}{\bibinfo{title}{The strength of friendship ties in
  proximity sensor data}}.
\newblock {\emph{\JournalTitle{PloS one}}} \textbf{\bibinfo{volume}{9}}
  (\bibinfo{year}{2014}).

\bibitem{luearly}
\bibinfo{author}{Lu, J.} \emph{et~al.}
\newblock \bibinfo{journal}{\bibinfo{title}{{COVID-19 Outbreak Associated with
  Air Conditioning in Restaurant, Guangzhou, China, 2020}}}.
\newblock {\emph{\JournalTitle{Emerg Infect Dis.}}}  (\bibinfo{year}{2020}).

\bibitem{landau2020}
\bibinfo{author}{Landau, S.}, \bibinfo{author}{Lopez, C.~E.} \&
  \bibinfo{author}{Moy, L.}
\newblock \bibinfo{journal}{\bibinfo{title}{{The Importance of Equity in
  Contact Tracing}}}.
\newblock {\emph{\JournalTitle{{Lawfare Blog}}}}  (\bibinfo{year}{2020}).

\bibitem{markup2020}
\bibinfo{author}{Lecher, C.}, \bibinfo{author}{Varner, M.} \&
  \bibinfo{author}{Martinez, E.}
\newblock \bibinfo{journal}{\bibinfo{title}{{Can You Get a Coronavirus Test? It
  May Depend on Where You Live}}}.
\newblock {\emph{\JournalTitle{{The Markup}}}}  (\bibinfo{year}{2020}).

\bibitem{article19}
\bibinfo{author}{{Multiple signatories [open letter]}}.
\newblock \bibinfo{journal}{\bibinfo{title}{{Coronavirus: States use of digital
  surveillance technologies to fight pandemic must respect human rights}}}.
\newblock {\emph{\JournalTitle{{Article19}}}}  (\bibinfo{year}{2020}).

\bibitem{osborne2020}
\bibinfo{author}{Osborne, C.} \& \bibinfo{author}{Day, Z.}
\newblock \bibinfo{journal}{\bibinfo{title}{{Proposed government coronavirus
  tracking app falls at the first hurdle due to data breach}}}.
\newblock {\emph{\JournalTitle{{ZDNet}}}}  (\bibinfo{year}{2020}).

\bibitem{wapo2020}
\bibinfo{author}{{Washington Post-University of Maryland}}.
\newblock \bibinfo{journal}{\bibinfo{title}{{National Poll, April 21-26,
  2020}}}.
\newblock {\emph{\JournalTitle{{The Washington Post}}}}
  (\bibinfo{year}{2020}).

\bibitem{chong2020}
\bibinfo{author}{Chong, C.}
\newblock \bibinfo{journal}{\bibinfo{title}{{About 1 million people have
  downloaded TraceTogether app, but more need to do so for it to be effective:
  Lawrence Wong}}}.
\newblock {\emph{\JournalTitle{{The Straits Times}}}}  (\bibinfo{year}{2020}).

\bibitem{olsen2020}
\bibinfo{author}{Olsen, S.~J.}
\newblock \bibinfo{journal}{\bibinfo{title}{{Smittestopp p{\aa} iPhone fungerer
  bare skikkelig n{\aa}r appen er {\aa}pen}}}.
\newblock {\emph{\JournalTitle{{Tek.no}}}}  (\bibinfo{year}{2020}).

\bibitem{bbc2020}
\bibinfo{author}{{BBC News}}.
\newblock \bibinfo{journal}{\bibinfo{title}{{Coronavirus lockdown protest:
  What's behind the US demonstrations?}}}
\newblock {\emph{\JournalTitle{{BBC}}}}  (\bibinfo{year}{2020}).

\bibitem{fa2020}
\bibinfo{author}{{Frakfurter Allgemeine}}.
\newblock \bibinfo{journal}{\bibinfo{title}{{Proteste in mehreren St\"adten}}}.
\newblock {\emph{\JournalTitle{{FAZ.NET}}}}  (\bibinfo{year}{2020}).

\bibitem{ettinger2016}
\bibinfo{author}{{Ettinger, Yair}}.
\newblock \bibinfo{journal}{\bibinfo{title}{{Ultra-Orthodox Jews Must Choose
  Between Obedience to Their Rabbis – or to Their Smartphone}}}.
\newblock {\emph{\JournalTitle{{Haaretz}}}}  (\bibinfo{year}{2016}).

\bibitem{hrc2020}
\bibinfo{author}{Kaye, D.}
\newblock \bibinfo{journal}{\bibinfo{title}{Disease pandemics and the freedom
  of opinion and expression}}.
\newblock {\emph{\JournalTitle{{A/HRC/44/49, para 54--57)}}}}
  (\bibinfo{year}{2020}).

\bibitem{reventlow2020}
\bibinfo{author}{Reventlow, N.~J.}
\newblock \bibinfo{journal}{\bibinfo{title}{{Why COVID-19 is a Crisis for
  Digital Rights}}}.
\newblock {\emph{\JournalTitle{{Digital Freedom Fund)}}}}
  (\bibinfo{year}{2020}).

\bibitem{phartiyal2020}
\bibinfo{author}{Phartiyal, S.}
\newblock \bibinfo{journal}{\bibinfo{title}{{Ahead of the curve: South Korea's
  evolving strategy to prevent a coronavirus resurgence}}}.
\newblock {\emph{\JournalTitle{{Reuters)}}}}  (\bibinfo{year}{2020}).

\bibitem{nowosielska2020}
\bibinfo{author}{Nowosielska, P.}
\newblock \bibinfo{journal}{\bibinfo{title}{{Luka w aplikacji Kwarantanna
  domowa. Kto poniesie za ni\k{a} konsekwencje?}}}
\newblock {\emph{\JournalTitle{{Dziennik Gazeta Prawna)}}}}
  (\bibinfo{year}{2020}).

\bibitem{tham2020}
\bibinfo{author}{Tham, I.}
\newblock \bibinfo{journal}{\bibinfo{title}{{No other way but to make use of
  TraceTogether mandatory}}}.
\newblock {\emph{\JournalTitle{{The Straits Times)}}}}  (\bibinfo{year}{2020}).

\bibitem{smith2020}
\bibinfo{author}{Smith, J.}, \bibinfo{author}{Hyonhee, S.} \&
  \bibinfo{author}{Cha, S.}
\newblock \bibinfo{journal}{\bibinfo{title}{{Ahead of the curve: South Korea's
  evolving strategy to prevent a coronavirus resurgence}}}.
\newblock {\emph{\JournalTitle{{Reuters)}}}}  (\bibinfo{year}{2020}).

\bibitem{zaik2020}
\bibinfo{author}{\v{Z}ajyk, D.}
\newblock \bibinfo{journal}{\bibinfo{title}{{Virusnyj monitoring: Kak
  koronavirus rasprostranil mirovuju sle\v{z}ku}}}.
\newblock {\emph{\JournalTitle{{The Steppe}}}}  (\bibinfo{year}{2020}).

\bibitem{nesterova2020}
\bibinfo{author}{Nesterova, E.}
\newblock \bibinfo{journal}{\bibinfo{title}{{Moskvi\v{c}ej s koronavirusom
  objazali prisylat` selfi - \v{c}toby podtverdit`, \v{c}to oni doma}}}.
\newblock {\emph{\JournalTitle{{Current Time}}}}  (\bibinfo{year}{2020}).

\bibitem{frank2020}
\bibinfo{author}{Frank, S.}
\newblock \bibinfo{journal}{\bibinfo{title}{{Golfstaaten setzen in Corona-Krise
  auf totale elektronische \"{U}berwachung}}}.
\newblock {\emph{\JournalTitle{{mena-watch}}}}  (\bibinfo{year}{2020}).

\bibitem{dixit2020b}
\bibinfo{author}{Dixit, P.}
\newblock \bibinfo{journal}{\bibinfo{title}{{An Entire City Has Been Told To
  Download A Controversial Contact Tracing App — Or Face Jail}}}.
\newblock {\emph{\JournalTitle{{BuzzFeed News}}}}  (\bibinfo{year}{2020}).

\bibitem{govpl2020}
\bibinfo{author}{{Ministerstwo Cyfryzacji}}.
\newblock \bibinfo{journal}{\bibinfo{title}{{Aplikacja Kwarantanna domowa –
  od dzi\'{s} obowi\k{a}zkowa}}}.
\newblock {\emph{\JournalTitle{{Serwis Rzeczypospolitej Polskiej}}}}
  (\bibinfo{year}{2020}).

\bibitem{marohovskaja2020}
\bibinfo{author}{Marohovskaja, A.} \& \bibinfo{author}{Velikovskij, D.}
\newblock \bibinfo{journal}{\bibinfo{title}{{Antisocial'nyj monitoring}}}.
\newblock {\emph{\JournalTitle{{.coda}}}}  (\bibinfo{year}{2020}).

\bibitem{southall2020}
\bibinfo{author}{Ashley, S.}
\newblock \bibinfo{journal}{\bibinfo{title}{{Scrutiny of Social-Distance
  Policing as 35 of 40 Arrested Are Black}}}.
\newblock {\emph{\JournalTitle{{The New York Times}}}}  (\bibinfo{year}{2020}).

\bibitem{olewe2020}
\bibinfo{author}{Olewe, D.}
\newblock \bibinfo{journal}{\bibinfo{title}{{Coronavirus in Africa: Whipping,
  shooting and snooping}}}.
\newblock {\emph{\JournalTitle{{BBC}}}}  (\bibinfo{year}{2020}).

\bibitem{greenberg2020}
\bibinfo{author}{Greenberg, A.}
\newblock \bibinfo{journal}{\bibinfo{title}{{Amid Its Covid-19 Crisis, China
  Was Still Hacking Uighurs’ iPhones}}}.
\newblock {\emph{\JournalTitle{{WIRED}}}}  (\bibinfo{year}{2020}).

\bibitem{gan2020}
\bibinfo{author}{Gan, N.}
\newblock \bibinfo{journal}{\bibinfo{title}{{China is installing surveillance
  cameras outside people's front doors ... and sometimes inside their homes}}}.
\newblock {\emph{\JournalTitle{{CNN Business}}}}  (\bibinfo{year}{2020}).

\bibitem{hasson2020}
\bibinfo{author}{Hasson, N.}
\newblock \bibinfo{journal}{\bibinfo{title}{{Amid Coronavirus Crisis, Israel
  Tells Palestinians to Download App That Tracks Phones }}}.
\newblock {\emph{\JournalTitle{{Haaretz}}}}  (\bibinfo{year}{2020}).

\bibitem{vedomosti2020}
\bibinfo{author}{Vedomosti}.
\newblock \bibinfo{journal}{\bibinfo{title}{{Rosgvardija na\v{c}ala
  ispol'zovat' bespilotniki dlja vyjavlenija naru\v{s}itelej samoizoljacii }}}.
\newblock {\emph{\JournalTitle{{Vedomosti.ru}}}}  (\bibinfo{year}{2020}).

\bibitem{vincent2020}
\bibinfo{author}{Vincent, J.}
\newblock \bibinfo{journal}{\bibinfo{title}{{France is using AI to check
  whether people are wearing masks on public transport}}}.
\newblock {\emph{\JournalTitle{{The Verge}}}}  (\bibinfo{year}{2020}).

\bibitem{rose2020}
\bibinfo{author}{Rose, J.}
\newblock \bibinfo{journal}{\bibinfo{title}{{Senate Votes to Allow FBI to Look
  at Your Web Browsing History Without a Warrant}}}.
\newblock {\emph{\JournalTitle{{Motherboard}}}}  (\bibinfo{year}{2020}).

\bibitem{bbc2020wrist}
\bibinfo{author}{{BBC News}}.
\newblock \bibinfo{journal}{\bibinfo{title}{{Coronavirus: People-tracking
  wristbands tested to enforce lockdown}}}.
\newblock {\emph{\JournalTitle{{BBC}}}}  (\bibinfo{year}{2020}).

\bibitem{mcdonald2020}
\bibinfo{author}{McDonald, S.}
\newblock \bibinfo{journal}{\bibinfo{title}{{Contact-tracing apps are
  political}}}.
\newblock {\emph{\JournalTitle{{Brookings - TechStream}}}}
  (\bibinfo{year}{2020}).

\bibitem{bbc2014}
\bibinfo{author}{{BBC News}}.
\newblock \bibinfo{journal}{\bibinfo{title}{{Edward Snowden: Leaks that exposed
  US spy programme}}}.
\newblock {\emph{\JournalTitle{{BBC}}}}  (\bibinfo{year}{2014}).

\bibitem{allen2020roadmap}
\bibinfo{author}{Allen, D.}, \bibinfo{author}{Block, S.},
  \bibinfo{author}{Cohen, J.} \emph{et~al.}
\newblock \bibinfo{journal}{\bibinfo{title}{Roadmap to pandemic resilience:
  Massive scale testing, tracing, and supported isolation (ttsi) as the path to
  pandemic resilience for a free society.}}
\newblock {\emph{\JournalTitle{Safra Center for Ethics at Harvard University}}}
  \textbf{\bibinfo{volume}{20}} (\bibinfo{year}{2020}).

\bibitem{dashboard2019}
\bibinfo{author}{{Center for Systems and Engineering (CSSE)}}.
\newblock \bibinfo{journal}{\bibinfo{title}{{COVID-19 Dashboard}}}.
\newblock {\emph{\JournalTitle{John Hopkins University}}}
  (\bibinfo{year}{2020}).

\bibitem{hoffower2020}
\bibinfo{author}{Hoffower, H.}
\newblock \bibinfo{journal}{\bibinfo{title}{{Rich urbanites are fleeing big
  cities and draining resources in smaller, more remote vacation spots. Here's
  where they're going — and how the locals feel about it.}}}
\newblock {\emph{\JournalTitle{Business Insider}}}  (\bibinfo{year}{2020}).

\bibitem{world2020considerations}
\bibinfo{author}{{World Health Organization and others}}.
\newblock \bibinfo{title}{{Considerations for quarantine of individuals in the
  context of containment for coronavirus disease (COVID-19): interim guidance,
  29 February 2020}}.
\newblock \bibinfo{type}{Tech. Rep.}, \bibinfo{institution}{World Health
  Organization} (\bibinfo{year}{2020}).

\bibitem{ilo2020}
\bibinfo{author}{{International Labour Organization}}.
\newblock \bibinfo{title}{{Contagion or starvation, the dilemma facing informal
  workers during the COVID-19 pandemic}}.
\newblock \bibinfo{type}{Tech. Rep.}, \bibinfo{institution}{{International
  Labour Organization}} (\bibinfo{year}{2020}).

\bibitem{bauer2020}
\bibinfo{author}{Bauer, L.}
\newblock \bibinfo{journal}{\bibinfo{title}{{The COVID-19 crisis has already
  left too many children hungry in America}}}.
\newblock {\emph{\JournalTitle{Brookings}}}  (\bibinfo{year}{2020}).

\bibitem{cevik2020}
\bibinfo{author}{{Dr Muge Cevik (@mugecevik)}}.
\newblock \bibinfo{title}{{[Tweet],
  {\url{https://twitter.com/mugecevik/status/1257392347010215947}}}}
  (\bibinfo{year}{2020}).

\bibitem{unstats2019}
\bibinfo{author}{{Department of Economic and Social Affairs, Statistics
  Division}}.
\newblock \bibinfo{title}{{Make cities and human settlements inclusive, safe,
  resilient and sustainable}}.
\newblock \bibinfo{type}{Tech. Rep.}, \bibinfo{institution}{United Nations}
  (\bibinfo{year}{2020}).

\bibitem{bird2017life}
\bibinfo{author}{Bird, J.}, \bibinfo{author}{Montebruno, P.} \&
  \bibinfo{author}{Regan, T.}
\newblock \bibinfo{journal}{\bibinfo{title}{Life in a slum: understanding
  living conditions in nairobi’s slums across time and space}}.
\newblock {\emph{\JournalTitle{Oxford Review of Economic Policy}}}
  \textbf{\bibinfo{volume}{33}}, \bibinfo{pages}{496--520}
  (\bibinfo{year}{2017}).

\bibitem{who2019}
\bibinfo{author}{Osseiran, N.} \& \bibinfo{author}{Lufadeju, Y.}
\newblock \bibinfo{title}{{1 in 3 people globally do not have access to safe
  drinking water – UNICEF, WHO}}.
\newblock \bibinfo{type}{Tech. Rep.}, \bibinfo{institution}{World Health
  Organization} (\bibinfo{year}{2019}).

\bibitem{walker2020}
\bibinfo{author}{Walker, S.}
\newblock \bibinfo{journal}{\bibinfo{title}{{Europe's marginalised Roma people
  hit hard by coronavirus }}}.
\newblock {\emph{\JournalTitle{The Guardian}}}  (\bibinfo{year}{2020}).

\bibitem{hadavas2020}
\bibinfo{author}{Hadavas, C.}
\newblock \bibinfo{journal}{\bibinfo{title}{{Singapore Contained the
  Coronavirus for Months. Now It Has One of the Worst Outbreaks in Asia.}}}
\newblock {\emph{\JournalTitle{Slate}}}  (\bibinfo{year}{2020}).

\bibitem{subbaraman2020distancing}
\bibinfo{author}{Subbaraman, N.}
\newblock \bibinfo{journal}{\bibinfo{title}{{'Distancing is impossible':
  refugee camps race to avert coronavirus catastrophe}}}.
\newblock {\emph{\JournalTitle{Nature}}}  (\bibinfo{year}{2020}).

\bibitem{marshall2020}
\bibinfo{author}{{The Marshall Project}}.
\newblock \bibinfo{title}{{A State-by-State Look at Coronavirus in Prisons}}.
\newblock \bibinfo{type}{Tech. Rep.}, \bibinfo{institution}{{The Marshall
  Project}} (\bibinfo{year}{2020}).

\bibitem{gurses2020}
\bibinfo{author}{{The Radical AI Podcast}}.
\newblock \bibinfo{title}{{Apple \& Google Partner to Promote Coronavirus
  Contact Tracing. Should You be Worried? with Seda Gurses}}.
\newblock
  \bibinfo{howpublished}{\url{https://podcasts.apple.com/us/podcast/apple-google-partner-to-promote-coronavirus-contact/id1505229145?i=1000471553068}}
  (\bibinfo{year}{2020}).

\bibitem{salathe2020}
\bibinfo{author}{Salath\'e, M.} \& \bibinfo{author}{Case, N.}
\newblock \bibinfo{title}{{ What Happens Next? COVID-19 Futures, Explained With
  Playable Simulations}}.
\newblock \bibinfo{howpublished}{\url{https://ncase.me/covid-19/}}
  (\bibinfo{year}{2020}).

\bibitem{georgia2020}
\bibinfo{author}{Lomsadze, G.}
\newblock \bibinfo{journal}{\bibinfo{title}{Georgia gets rare plaudits for
  coronavirus response}}.
\newblock {\emph{\JournalTitle{{eurasianet}}}}  (\bibinfo{year}{2020}).

\bibitem{ghana2020}
\bibinfo{author}{Hirsch, A.}
\newblock \bibinfo{journal}{\bibinfo{title}{{Why are Africa's coronavirus
  successes being overlooked?}}}
\newblock {\emph{\JournalTitle{{The Guardian}}}}  (\bibinfo{year}{2020}).

\bibitem{diplomat2020}
\bibinfo{author}{Nowrojee, B.}
\newblock \bibinfo{journal}{\bibinfo{title}{{How a South Indian State Flattened
  Its Coronavirus Curve}}}.
\newblock {\emph{\JournalTitle{{The Diplomat}}}}  (\bibinfo{year}{2020}).

\bibitem{duffin2020}
\bibinfo{author}{Simmons-Duffin, S.}
\newblock \bibinfo{journal}{\bibinfo{title}{{States Nearly Doubled Plans For
  Contact Tracers Since NPR Surveyed Them 10 Days Ago}}}.
\newblock {\emph{\JournalTitle{{NPR}}}}  (\bibinfo{year}{2020}).

\end{thebibliography}

\end{document}